\documentclass[12pt,dvips]{article}
\usepackage{amssymb, latexsym, amsmath, epsfig, subfigure}

\begin{document}

\title{Dynamics of radiation dominated branes: 
Vacuum dynamics from radiation}

\author{Peter D. Rippis\thanks{e-mail p.d.rippis@fys.uio.no}\\
Center for Advanced Study\\
Drammensveien 78, Oslo Norway}

\date{May 30, 2005}

\maketitle

\begin{abstract}
I point out a symmetry, between equations of state for polytropic fluids, 
in the equation of motion of a spherically symmetric singular shell 
embedded in  4-d and 5-d vacuum spacetimes. 
In particular the equation of motion of a 
shell consisting of radiation has the same form as  
for a vacuum shell or domain wall.  
\end{abstract}

\section{Introduction}

In general relativity geometry and energy are non-linearly connected, 
and to study its dynamics is in general very complicated. 
Therefore we introduce 
symmetries. I shall consider  spherically symmetric static $d+1$ dimensional 
spacetimes, $\mathcal{M}$, for $d=3$ and $d=4$; this could be Minkowski, 
Schwarzschild, de-Sitter and anti de-Sitter spacetimes.  
In these spacetimes we embed matter or energy confined to a $d$ 
dimensional surface or hypersurface (also denoted as a d-1 brane, 
referring to the number of spatial dimensions), 
i.e. the length scales are such 
that the thickness of the surface can be ignored. And the equation of motion 
is reduced to the 1-dimensional evolution of the radius of 
the surface. 

A very useful formalism to study 
such singular surfaces in general relativity was introduced by Israel\cite{is} 
and has been extensively studied  
both for static and non-static spacetimes, see  \cite{grorip} and 
references therein. 

The surface does not in general follow the geodesics of the 
background or bulk spacetime. This gives rise to some interesting results, 
e.g radiation, material and vacuum shells 
that have no gravitational field.
Radiation (photonic) shells have been studied in a 
cylindrical spacetime \cite{bon,ardel}.
In a 4-dimensional spherical spacetime of pure vacuum ($\Lambda =0$) 
we show that for a radiation shell we get 
the potential of a domain wall \cite{bgg} and discuss the solutions.

Recent observations \cite{ri,per} indicate that the expansion of the
universe is accelerating. Indications for this was also found in 1966
\cite{sol}. Motivated by string theory one 
considers 5-dimensional spacetimes where a 
singular shell or brane represents the universe. (See \cite{lang} for a 
review on brane cosmology.) 
The acceleration of the universe could be explained by  a form of dark energy. 
This can be described in three ways: by a cosmological constant $\Lambda$,
an ultralight scalar field or by a modification of Einstein's equations, e.g.
\cite{kmp} where a radiation dominated brane is found to have accelerated 
expansion. The main new point of the present paper is that electro-magnetic 
radiation  can contribute to $\Lambda$ without any corrections to the 
field equations. Here I do not invoke the $Z_{2}$ symmetry.

\section{Equation of motion}

To study singular shells or branes in general relativity we use the 
metric junction method of Israel \cite{is}. 
Hence, the spacetime manifold, $\mathcal{M}$, is split into two parts  
$\mathcal{M}^{+}$ and $\mathcal{M}^{-}$ separated by a common boundary 
$\Sigma$. 
The Einstein field equations are
\begin{equation}\label{eq:Efe}
R_{\mu\nu}-\frac{1}{2}Rg_{\mu\nu}=-\Lambda g_{\mu\nu} + 
\kappa^{2}T_{\mu\nu},
\end{equation}
$\kappa^{2}$ is the  gravitational coupling constant in $d+1$ dimensions.
In the 4-dimensional spacetime of genral relativity we have
$\kappa^{2}=\frac{8\pi\mathcal{G}}{c^{4}}$.
We use greek indices to 
denote the $d+1$ coordinates of the bulk $\mathcal{M}^{\pm}$  
and latin indices for the $d$ coordinates on the hypersurface $\Sigma$.

The equation of motion for the hypersurface is found from evaluating
the Einstein field equations (\ref{eq:Efe}) across the boundary and
is contained in the Lanczos equation, which for a 
$d+1$ dimensional spacetime is
\begin{equation}\label{eq:lancz}
[K_{ij}]=\kappa^{2}\Big(S_{ij}-\frac{1}{d-1}Sg_{ij}\Big),
\end{equation}
where $[K_{ij}]$ is the discontinuity of the extrinsic curvature tensor at 
the surface. $K_{ij}$ is defined as the covariant derivative of 
the normal vector ${\bf n}$ to the hypersurface: 
$K_{ij}=-{\bf e}_{i}\cdot\nabla_{j} {\bf n}$.
$S_{ij}$ is the energy-momentum tensor for the surface. 
For an ideal fluid $S_{ij}$ is given by
\begin{equation}\label{eq:if}
S_{ij}=(\rho+p)u_{i}u_{j} +p\;g_{ij},
\end{equation}
where $\rho$ is the mass (energy) density of the surface
 and $p$ is the tangential pressure of
the surface. For a radiation dominated brane we 
have $S_{i}^{~i}=0$. Thus, for radiation $p=\frac{1}{d-1}\rho$.

The continuity equation for the brane is
\begin{equation}\label{eq:cont}
S_{i~|j}^{~j}+[T_{i\mu}n^{\mu}]=0, 
\end{equation}
where $|$ denotes covariant derivative with respect to the metric connection on the brane.
For a vacuum bulk we have $[T_{i\mu}n^{\mu}]=0$, i.e. there are no normal forces 
exerted on the brane by the surroundings. 
We also have:
\begin{eqnarray}
S_{ij}[K^{ij}]&=& 
\kappa^{2}\left(S_{ij}S^{ij}-\frac{1}{d-1}S^{2}\right) \\ 
S_{ij}\{ K^{ij}\}&=&-[T_{\mu\nu}n^{\mu}n^{\nu}]. 
\end{eqnarray}

\section{4-dimensional spacetimes}
For spherically symmetric static 4-dimensional spacetimes the metric 
in $\mathcal{M}^{\pm}$ can be written
\begin{equation}
ds^{2}_{\pm}=-f_{\pm}dT^{2}_{\pm}+f_{\pm}^{-1}dR^{2}+R^{2}d\Omega^{2}.
\end{equation}
We shall look at a shell embedded in vacuum so that outside the shell we have
Schwarzschild/de-Sitter spacetimes: 
\begin{equation}
f_{\pm}=1-\frac{2m_{\pm}}{R}-
\frac{\Lambda_{\pm}}{3}R^{2}.
\end{equation}
The Schwarzschild mass $m_{-}$ gives a mass at the center and
$m_{+}=m_{-}+m$, with $m$ the Schwarzschild mass of the shell. 

For intrinsic coordinates  we use proper time, $\tau$, 
on the shell. Thus, the metric on the shell is
\begin{equation}
ds^{2}_{\Sigma}=-d\tau^{2}+R^{2}d\Omega^{2}.
\end{equation}

The normal vector to the shell is
\begin{equation}
n_{\alpha}^{\pm}=(-\dot{R},\dot{T}_{\pm},0,0).
\end{equation}
From requiring that we induce the same metric on 
$\Sigma$ from $\mathcal{M}^{-}$ and $\mathcal{M}^{+}$ we have
$\dot{T}_{\pm}=f^{-1}_{\pm}\sqrt{f_{\pm}+\dot{R}^{2}}$, where 
$\dot{}\equiv\frac{d}{d\tau}$.

In this case the continuity equation (\ref{eq:cont}) for an ideal fluid 
(\ref{eq:if}) gives
\begin{equation}\label{eq:contif}
\dot{\rho}=-2(\rho+p)\frac{\dot{R}}{R}. 
\end{equation}
For an equation of state $p=\omega \rho$ with $\omega$  constant,
this is readily solved. Thus, $\rho$ in terms of the trajectories 
$R(\tau)$ of the shell is
\begin{equation}\label{eq:omcon}
 \rho=bR^{-2(1+\omega)},\quad b=constant. 
\end{equation}
For a dust shell, $\omega=0$, the constant $b$ gives a measure of the total 
rest mass of the shell.

The components of the extrinsic curvature tensor are:
$K^{\pm}_{\theta\theta}=\zeta_{\pm}R\sqrt{f_{_{\pm}}+\dot{R}^{2}}$ and
$K^{\pm}_{\tau\tau}=\frac{-\zeta_{\pm}}{\sqrt{f_{_{\pm}}+\dot{R}^{2}}}
(\ddot{R}+\frac{1}{2}\frac{\partial f_{\pm}}{\partial R})$,
where $\zeta=\pm1$. For static spacetimes 
the sign given by $\zeta$ gives the spatial topology\footnote{With our sign 
convention the area of the shell is increasing in the normal direction 
for $\zeta=-1$ and decreasing for $\zeta=+1$. 
$\zeta$ can also be identified with the angle 
$arctan{\frac{v}{u}}$ in Kruskal-Szekers type coordinates, see \cite{bgg}; 
for $\zeta=-1$
the angle is increasing along the shells trajectory and for
$\zeta=+1$  the angel is decreasing along the trajectory.}.
The sign $\zeta_{\pm}$ can be found from squaring the angular component of 
(\ref{eq:lancz}):
\begin{equation}\label{eq:sign}
K_{\theta\theta}^{\pm}= \frac{1}{8\pi \rho}(f_{+}-f_{-})\pm 2\pi R^{2}\rho
\end{equation}
where the $\pm$ superscript corresponds to the $\pm$ in the equation 
respectively.
An expression for the acceleration not containing discontinuities or averages
is given by \cite{grorip}
\begin{eqnarray}\label{eq:acc}
K_{\tau\tau}^{\pm}=-\frac{[T_{\mu\nu}n^{\mu}n^{\nu}]}{\rho}-
\frac{2p}{R^{2}\rho}K_{\theta\theta}^{\pm}\pm \frac{1}{2}\kappa^{2}
\left(\frac{1}{2}\rho+2p\right),
\end{eqnarray}
where the normal component of the four acceleration is given by
$a^{\pm}\equiv a_{\mu}n^{\mu}|^{\pm}=K_{\tau\tau}^{\pm}$.
The first term gives the difference in the normal force on the shell
 from the bulk, and we have $[T_{\mu\nu}n^{\mu}n^{\nu}]=0$. 
The second term is due to the pressure of the shell and the 
geometry of the embedding. 
The last term gives the self gravitational attraction.
Thus, from (\ref{eq:acc}) and (\ref{eq:sign}) we have
for $\Lambda_{-}=\Lambda_{+}=\Lambda$:
\begin{equation}\label{eq:nacc}
a^{\pm}= -\frac{2\omega}{R^{2}}\left(-\frac{m}{4\pi R\rho}\pm
2\pi R^{2}\rho \right) \pm 4\pi\rho\left(\frac{1}{2}+2\omega\right),
\end{equation}
independent of $m_{-}$ and $\Lambda$. 
From (\ref{eq:sign}) we find that
$\zeta_{-}=-1$ for all $R$, while $\zeta_{+}$ will change sign during the 
history of the shell. ($\zeta_{-}$ changes sign if $\Lambda_{-}>\Lambda_{+}$).

From the angular component of (\ref{eq:lancz}) the equation of motion for 
a surface consisting of an ideal fluid is:
\begin{equation}\label{eq:rdotgen}
\dot{R}^{2}=\frac{1}{(\kappa^{2} R\rho)^{2}}(f_{+}-f_{-})^{2}-
\frac{1}{2}(f_{+}+f_{-})+\frac{1}{2}(\kappa^{2} R\rho)^{2}.
\end{equation}
Taking the derivative of this and using the continuity equation 
(\ref{eq:contif}) gives the 
time component of the Lanczos equation (\ref{eq:lancz}), 
i.e. the dynamics is contained in (\ref{eq:rdotgen}).

From eq. (\ref{eq:rdotgen}) using  eq. (\ref{eq:omcon})  
we  get\footnote{We have set $\kappa^{2}=8\pi$, i.e. 
in geometrical units $c=1$ and $\mathcal{G}=1$} 
\begin{equation}\label{eq:sym4d}
\dot{R}^{2}=\Big(\frac{m}{4\pi b}\Big)^{2}R^{4\omega}+ 
(2\pi b)^{2}R^{-(4\omega+2)}+\frac{m+2m_{-}}{R}+\frac{\Lambda}{3}R^{2}-1.
\end{equation}
The first term comes from the embedding, the second term is from 
the gravitational self interaction of the shell, and the last terms are due 
to the background.

We observe an interesting symmetry in eq. (\ref{eq:sym4d}).
For equation of states given by $\omega_{1}$ and $\omega_{2}$ where
\begin{equation}\label{eq:omg4d12}
\omega_{2}=-\Big(\omega_{1}+\frac{1}{2}\Big)
\end{equation}
the equation of motion (\ref{eq:sym4d}) is invariant under the 
 transformation
$\omega_{1} \rightarrow \omega_{2}$ with the interchange
\begin{equation}
\frac{m}{4\pi b}\leftrightarrow 2\pi b.
\end{equation}
That is, a radiation shell $\omega=\frac{1}{2}$ is symmetric to vacuum shell 
(or domain wall) $\omega=-1$, while a dust shell, $\omega=0$, is symmetric 
to a shell with $\omega=-\frac{1}{2}$. 
For $\omega_{1}>\frac{1}{2}$ we have  $\omega_{2}<-1$, 
and for equation of state $-\frac{1}{4}$ we get $\omega_{1}=\omega_{2}$. 

Let us look at solutions of eq. (\ref{eq:sym4d}) with $\Lambda=0$ for 
small and large $R$.
For a radiation shell at small $R$ the motion 
is dictated by the second term, i.e.
$R\propto \tau^{\frac{1}{3}}$. For large $R$ the first term dominates giving 
exponential evolution, $ln{R}\propto \tau$.
A dust shell at small $R$ evolves with  $R\propto \sqrt{\tau}$,
while at large $R$ we find $R\propto \tau$.

\subsection{Radiation shells}

For a radiation shell in a Schwarzschild/de-Sitter background we get:
\begin{equation}
\dot{R}^{2}=\alpha R^{2}+ \frac{(2\pi b)^{2}}{R^{4}}+\frac{m+2m_{-}}{R}-1,
\end{equation}
where
\begin{equation}
\alpha=\frac{m^{2}}{(4\pi b)^{2}}+\frac{\Lambda}{3}.
\end{equation} 
The acceleration can be written from (\ref{eq:nacc})
\begin{equation}\label{eq:arad}
a^{\pm}=\frac{m}{4\pi b}\pm \frac{4\pi b}{R^{3}},
\end{equation}  
we see that $a^{-}$ changes sign at
$R^{3}=\frac{(4\pi b)^{2}}{m}$. 

The total constant mass, $m$, of the shell
can be written
\begin{eqnarray}
m&=&\frac{4\pi b}{R}\sqrt{f_{-}+\dot{R}^{2}}-
     \frac{(4\pi b)^{2}}{2R^{3}} \nonumber \\
&=&\frac{4\pi b}{2R}\left(\zeta_{+}\sqrt{f_{+}+\dot{R}^{2}}+
\zeta_{-}\sqrt{f_{-}+\dot{R}^{2}}\right).
\end{eqnarray}  
If $\zeta_{+}=+1$ ($\zeta_{-}=-1$) throughout the history of the shell
the equation of motion 
allows for a shell with zero gravitational mass, i.e. the shell is 
embedded in a flat spacetime.

\begin{figure}[!t]
\begin{center}
\scalebox{0.9}{
\includegraphics{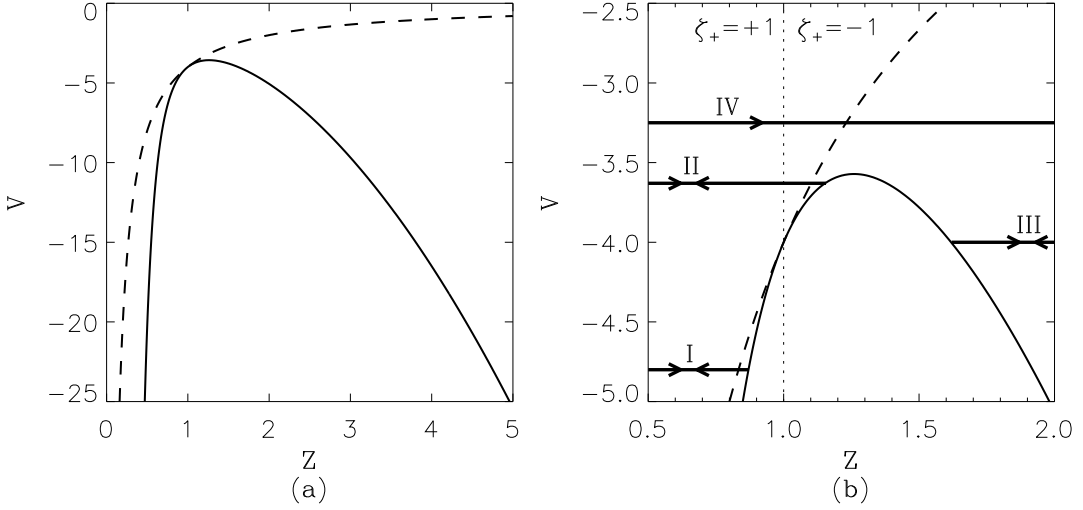}}
\caption{Graph (a) shows the potential $V(Z)$
for a radiation dominated thin shell. 
The dashed line gives the event horizon associated
with this potential. In (b) we show  the 4 different solutions. 
The dotted line is where the potential and the horizon are equal and where the 
sign $\zeta_{+}$ changes sign.}\label{fig:figvacbub} 
\end{center}
\end{figure}

Let us consider the case where $\Lambda=0$ and $m_{-}=0$. 
The equation of motion for the 
shell can also be viewed as describing a particle in a 1-dimensional potential,
and introducing the variables:
\begin{eqnarray}\label{eq:zrad}
Z^{3}&=&\frac{2m}{(4\pi b)^{2}}R^{3}\\
t&=&\frac{m}{4\pi b}\tau,
\end{eqnarray}
we obtain the same potential as in \cite{bgg} for a domain wall\footnote{
In their notation  the potential is for 
$\gamma=2$ which is for $\chi=0$, i.e. $\Lambda_{\pm}=0$.}:
\begin{equation}
\left(\frac{dZ}{dt}\right)^{2} +V = E,
\end{equation}
where
\begin{eqnarray}
V&=&-\left(\frac{Z^{3}-1}{Z^{2}}\right)^{2}-
\frac{4}{Z}\\ \label{eq:E}
E&=&\frac{-4(4\pi b)^{\frac{2}{3}}}{(2m)^{\frac{4}{3}}}.
\end{eqnarray}
The Schwarzschild horizon is given by $R_{H}=2m$ which from  
eq. (\ref{eq:zrad}) and eq. (\ref{eq:E})
leads to $E(Z_{H})=-\frac{4}{Z_{H}}$.
The potential is shown in fig. (\ref{fig:figvacbub}).
For all the details on this potential see 
\cite{bgg}.

We have seen that $\zeta_{-}=-1$ for all $R$, 
while for $\zeta_{+}$ we find changes from $+1$ to $-1$
at $Z_{+}=1$ for increasing $Z$.\footnote{For a domain wall it is opposite, 
$\zeta_{+}$ goes from $-1$ to $+1$.} 
Also, the mass for this energy level
$V(Z_{+})=E(m_{+})$ is
\begin{equation}
m_{+}=(\pi b)^{\frac{1}{2}},
\end{equation}
i.e. if if we have  $m>m_{+}$ then $\zeta_{+}$ changes sign during the motion 
of the shell.

The maximum of the potential is at $Z_{m}^{3}=2$, see (\ref{eq:arad}) 
and (\ref{eq:zrad}), 
and is found to be \cite{bgg} $V(Z_{m})=-\frac{9}{2^{\frac{4}{3}}}$.
The critical mass for unbound motion, $V(Z_{m})=E(m_{cr})$, is
\begin{equation}
m_{cr}=\Big(\frac{32}{27}\pi b\Big)^{\frac{1}{2}}
\approx 1.09(\pi b)^{\frac{1}{2}}.
\end{equation}
For $m>m_{cr}$ we have unbound motion, solution $IV$ in 
fig.\ref{fig:figvacbub}. In all we have four solutions:\\
\\
Solution $I$: This is a bound solution for a mass  $m<m_{+}$. 
So that $\zeta_{+}=+1$
throughout the history of the shell. The shell expands from $R=0$ to a maximum
$R$ outside the Schwarzschild horizon and then contracts, 
see fig. 8 (b) in \cite{bgg}. Also, this solution admits a zero gravitational mass, $m=0$.\\
\\
Solution $II$: This is also a bound solution, with $m$ in the range 
$m_{+}<m<m_{cr}$. 
So that now $\zeta_{+}=-1$ when the shell crosses the horizon.
See fig. 7 (b) in \cite{bgg}.\\
\\
Solution $III$: This is a bounce solution for $m<m_{cr}$.
Here $\zeta_{+}=-1$ throughout the history of the shell.
 The shell comes in from infinity stops at a 
minimum $R$ outside the Schwarzschild horizon and heads out again. This 
corresponds to fig. 10 (b) in \cite{bgg} with the shell in region $I$ 
instead of region $II$.\\
\\
Solution $IV$: This is an unbound solution, where $m>m_{cr}$. 
Here $\zeta_{+}=-1$ when the shell crosses the horizon. This
corresponds to fig. 11 (b) in \cite{bgg} with the shell moving into region 
$I$ instead of region $II$.\\

Consider now a shell consisting of radiation and dust,
$\rho=\rho_{r}+\rho_{d}$. The radiation density goes as $R^{-3}$ while the dust density goes with $R^{-2}$. Hence, for small $R$ the motion will be dictated by the radiation, and for large $R$ the motion is determined by the dust; i.e.
there is no exponential expansion for large $R$ as for a domain wall 
containing dust. 

\section{5-dimensional spacetimes} 
We now consider a brane cosmological model without the $Z_{2}$ symmetry. 
We look at a brane that represents the boundary between 
Schwarschild/(anti)de-Sitter spacetimes, see also \cite{bcg}. 
The 5-dimensional spacetime  metric  can be written 
\begin{equation}
ds^{2}=-FdT^{2}+F^{-1}dR^{2}+R^{2}d\Omega^{2}_{4}
\end{equation}
with
\begin{equation}
d\Omega^{2}_{4}=\frac{d\chi^{2}}{1-k\chi^{2}}+
sin^{2}\chi(d\theta^{2}+sin^{2}\theta d\phi^{2})
\end{equation}
and
\begin{equation}
F_{\pm}=k_{\pm}-\frac{{\cal C}_{\pm}}{R^{2}}-\frac{\lambda_{\pm}}{6}R^{2},
\end{equation}
$\mathcal{C}$ is the 5-dimensional equivalent of the Schwarzschild mass and 
$\lambda$ is the cosmological constant of the bulk. 
$k=0,\pm1$ gives the spatial curvature, planar or spherical/hyperboloidal.

The intrinsic metric on the brane is
\begin{equation}
ds^{2}_{_{\Sigma}}=-d\tau^{2}+R^{2}d\Omega^{2}_{4},
\end{equation}
with $R=R(\tau)$ representing the expansion factor of the universe.

The continuity equation gives
\begin{equation}
\dot{\rho}=-3(\rho+p)\frac{\dot{R}}{R}
\end{equation}
with solution for a polytropic equation of state, $p=\omega\rho$, given by
\begin{equation}\label{eq:rho} 
\rho=bR^{-3(\omega+1)}.
\end{equation}

The spatial components of the extrinsic curvature tensor are up to a sign 
\begin{equation}
K_{ij}^{\pm}=\frac{\sqrt{f_{\pm}+\dot{R}^{2}}}{R}g_{ij}.
\end{equation}
The resulting Friedmann equation has the same  form as eq. (\ref{eq:rdotgen}):
\begin{equation}\label{eq:rdot5d}
\dot{R}^{2}=\frac{1}{(\frac{2}{3}\kappa^{2} R\rho)^{2}}(F_{+}-F_{-})^{2}-
\frac{1}{2}(F_{+}+F_{-})+(\frac{\kappa^{2}}{6} R\rho)^{2}.
\end{equation}
Here we have $\rho^{2}$ and $\rho^{-2}$ dependence. 
In $Z_{2}$ symmetric brane cosmology the first term vanishes, i.e. 
we only have the  $\rho^{2}$ dependence.

Inserting for
$\rho$ from eq. (\ref{eq:rho}) we arrive at
\begin{equation}\label{eq:sym5d}
\dot{R}^{2}=\Big(\frac{3\mathcal{C}}{2\kappa^{2} b}  \Big)^{2} R^{6\omega} +
            \Big(\frac{\kappa^{2} b}{6}  \Big)^{2}R^{-(6\omega+4)} +
            \frac{\Lambda}{6} R^{2} + \frac{\mathcal{C}}{2R^{2}} - k,
\end{equation}
with $k=k_{+}=k_{-}$ and $\lambda=\lambda_{-}=\lambda_{+}$, and
where we have set $\mathcal{C}_{-}=0$. Thus, this gives the equation of motion
for the boundary between (anti)de-Sitter or Minkowski and 
Schwarzschild/(anti)de-Sitter spacetimes.
Equation (\ref{eq:sym5d}) is invariant under the transformation
$\omega_{1} \rightarrow \omega_{2}$
where
\begin{equation}\label{eq:omg5d12}
\omega_{2}=-\Big(\omega_{1}+\frac{2}{3}\Big)
\end{equation}
with the interchange 
\begin{equation}
\frac{3\mathcal{C}}{2\kappa^{2} b}\leftrightarrow
\frac{\kappa^{2} b}{6}.
\end{equation}

Hence, radiation, 
$\omega=\frac{1}{3}$, corresponds to Lorentz invariant  vacuum energy, 
$\omega=-1$. 
For $\omega_{1}>\frac{1}{3}$ we get $\omega_{2}<-1$.
A dust universe, $\omega=0$, is symmetric to  
$\omega=-\frac{2}{3}$, which represent a topological defect in form of a 
domain wall. A cosmic string $\omega=-\frac{1}{3}$ has no symmetry, 
$\omega_{1}=\omega_{2}$.

\subsection{Radiation and dust branes}
For a radiation dominated universe the Friedmann equation becomes
\begin{equation}\label{eq:rad5d}
\dot{R}^{2}=\alpha R^{2} +
            \beta R^{-6} +
             + \frac{\mathcal{C}}{2R^{2}} - k
\end{equation}
with
\begin{equation}
\alpha=\left(\frac{3\mathcal{C}}{2\kappa^{2} b}\right)^{2}+\frac{\lambda}{6},
\quad \beta=\Big(\frac{\kappa^{2} b}{6} \Big)^{2}.
\end{equation}
For $k=0$, and defining $X=R^{4}$ this can be written
\begin{equation}
\frac{\dot{X}^{2}}{16}=\alpha X^{2} +\frac{\mathcal{C}}{2}X + \beta.
\end{equation}
This has same form as a $Z_{2}$ symmetric brane with radiation and 
vacuum energy (with $\mathcal{C}=0$), see e.g. \cite{lang}. 
Integrating (\ref{eq:rad5d}) with  
$\alpha=0$ we have the expansion $R^{4}=2\mathcal{C}t^{2}+4\sqrt{\beta}t$.
Thus, at late time we have standard evolution.   
But for a brane containing radiation and dust the equation of motion will
at late times be determined by the dust component.   
To consider  dust solutions  we set $\lambda=0$ and define $Y=R^{2}$ to find
\begin{equation}
\frac{\dot{Y}^{2}}{4}=\alpha' Y + \beta Y^{-1}+\frac{\mathcal{C}}{2}
\end{equation}
with
\begin{equation}
\alpha'=\left(\frac{3\mathcal{C}}{2\kappa^{2} b}\right)^{2}-k.
\end{equation}
This is solved by elliptical integrals. We only look at the limits.
For small $R$ we  get the evolution 
$R\propto \tau^{\frac{1}{3}}$, for large $R$ we get 
$R\propto \tau$, i.e. we do not get the standard evolution of 
$\tau^{\frac{2}{3}}$.

\section{Conclusions}
We have studied spherically symmetric branes embedded in static spacetimes, 
and shown that the equation of motion in  4-dimensional spacetimes, 
eq. (\ref{eq:sym4d}), and  5-dimensional spacetimes, eq. (\ref{eq:sym5d}), 
have the same form  for different polytropic fluids given 
by $\omega_{1}$ and $\omega_{2}$. Where $\omega_{1}$ and $\omega_{2}$ are 
related by eq. (\ref{eq:omg4d12}) and eq. (\ref{eq:omg5d12}) respectively. 
This symmetry comes from the embedding and is of second order in the 
Schwarzschild mass of the shell.
An interesting case is that a radiation shell corresponds to a 
vacuum shell or domain wall.

In 4 dimensions we investigate the equation of motion for a 
radiation shell embedded in an empty spacetime, i.e.
Minkowski and Schwarzschild spacetimes outside the shell. 
The potential describing the motion is
identical to that of a domain wall studied exhaustively in \cite{bgg}. 
We find different global properties.
Also, in contrast to a domain wall, for a radiation shell with added dust 
the dust component will dominate for large radii.    

In  5 dimensions we show that the equation of motion for a radiation brane 
has the same form as a mirror, $Z_{2}$,
 symmetric brane with vacuum enrgy and radiation.
Where for late times we find standard cosmological evolution.
For a dust brane we find that it evolves as  Randall-Sundrum for early times 
will at late times the evolution is proportional to cosmic time $\tau$. 

\section{Acknowledgment}
I wish to thank \O yvind Gr\o n for many interesting discussions and 
comments on an initial version of the manuscript. 
This work was done while affiliated with the Center for Advanced Study, 
Oslo Norway, in the project ``Turbulence in plasmas and fluids''
headed by Hans L. P\'{e}cseli and Jan Trulsen.


\begin{thebibliography}{99}


\bibitem{is} Israel, W.  
Il Nouvo Cimento  $\mathbf{44}$B, 1 (1966) (Erratum) $\mathbf{48}$B, 463.

\bibitem{grorip}\O. Gr\o n and P. D. Rippis. (2003)
Gen. Rel.  Grav.  $\mathbf{35}$, 2189 . Erratum (2004)
$\mathbf{36}$, 1505.  [gr-qc/0307006].

\bibitem{bon} Bonnor, W. B. (2005).
Class. Quantum Grav. $\mathbf{22}$, 803.

\bibitem{ardel} Arik, M. and Delice, O. (2003). Gen. Rel. Grav. 
$\mathbf{34}$, 1285.

\bibitem{bgg} Blau, S. K., Guendelman, E. I. and Guth, A. H. (1987).
 Phys. Rev. D $\mathbf{35}$, 1747.

\bibitem{ri} Riess, A. G., et al. (1998). AJ, $\mathbf{116}$, 1009

\bibitem{per} Perlmutter, S. et al. (1999). Ap. J, $\mathbf{517}$, 565

\bibitem{sol} Solheim, J.-E. (1966). MNRAS, $\mathbf{133}$, 32

\bibitem{lang} Langlois, D. (2003). Prog. Theor. Phys. Suppl.
 $\mathbf{148}$, 181, [hep-th/0209261].

\bibitem{kmp} Kofinas, G., Maartens, R. and Papantonopoulos, E. (2003).
JHEP  $\mathbf{0310}$, 066. [hep-th/0307138]. Papantonopoulos, E. (2004)
[gr-qc/0402115].

\bibitem{bcg} Bowcock, P., Charmousis, C. and Gregory, R (2000)
Class. Quant. Grav. $\mathbf{17}$ 4745, [hep-th/0007177].


\end{thebibliography}
\end{document}